\documentclass[aps,pra,twocolumn]{revtex4}

\usepackage{graphicx}
\usepackage{pstricks}
\usepackage{bm}

\newcommand{\mb}{\mathbf}
\newcommand{\be}{\begin{equation}}
\newcommand{\ee}{\end{equation}}
\newcommand{\bwt}{\begin{widetext}}
\newcommand{\ewt}{\end{widetext}}

\begin{document}

\title{Superradiant Raman scattering in an ultracold Bose gas at finite
temperature}
\author{H.~Uys\footnote{Currently at the Time and Frequency Division, National Institute of Standards and Technology,
Boulder, CO, 80305.} and P.~Meystre}
\affiliation{Department of Physics and B2 Institute,
The University of Arizona, Tucson, AZ, 85721}
\begin{abstract}
We study superradiant Raman scattering from an ultra-cold, but
finite temperature Bose gas in a harmonic trap. Numerical
simulations indicate the existence of distinct timescales
associated with the decoherence of the condensed versus thermal
fractions, and the concomitant preferred scattering from atoms in
low lying trap states in the regime where superradiance takes
place on a timescale comparable to an inverse trap frequency.  As a
consequence the scattered atoms experience a modest reduction in
temperature as compared to the unscattered atoms.
\end{abstract}
\maketitle

\section{Introduction}

Bose-Einstein condensation (BEC) is characterized by the appearance
off-diagonal long-range order, with phase correlations between
points spatially separated over macroscopic distances.  These
correlations are quantified by the first-order spatial correlation
function
$G^{(1)}(\mb{x},\mb{x}^\prime)=\langle\hat\psi^\dagger(\mb{x})\hat\psi(\mb{x}^\prime)\rangle$,
where $\hat\psi^\dagger(\mb{x})$ is the field operator that creates a
particle at position $\mb{x}$. In the thermodynamic limit
$G^{(1)}(\mb{x},\mb{x}^\prime)$ becomes finite-valued for infinitely
separated spatial points below the critical temperature. In
ultracold atomic systems the presence of long-range phase
correlations is experimentally manifest through the direct
observation of matter-wave interference patterns. Hence, matter-wave
interference has been used as a diagnostic tool to
demonstrate the appearance of the superfluid to Mott-insulator
transition \cite{Greiner2002T}, the growth of spatial correlations
during the formation of a Bose-condensate from a non-equilibrium
situation after sudden quenching across the transition point
\cite{Ritter2007T}, and to measure the critical exponent
characterizing the divergence of the correlation length near that
point \cite{Donner2007T}.

Another tool recently employed to probe the BEC transition is the
selective sensitivity of superradiant scattering to the condensed
fraction of an ultra-cold Bose gas beneath the critical point for
condensation \cite{Sadler2007T}. Early demonstrations of
superradiance were realized in thermal gases at temperatures in
the several hundreds of degrees range,
\cite{Skribanowitz1973T,Gross1976T,Gibbs1977T,Rosenberger1981T}.
The majority of these experiments involving initially inverted
atomic transitions in relatively high-density samples, a situation
often referred to as \textit{superfluorescence}
\cite{Bonifacio1975T,Rosenberger1981T}. By contrast, recent experiments
in ultra-cold atomic vapors involved off-resonant light scattering
in very low density samples. This regime was first studied in 1999
\cite{Inouye1999aT} and has since been the focus of several
experiments, such as in the context of coherent matter-wave
amplification \cite{Inouye1999bT,Kozuma1999T,Schneble2004T}. The
majority of these studies used almost pure condensates in which
superradiant enhancement is strong due to the near absence of
Doppler dephasing.  In addition to related  theoretical
\cite{Bonifacio1997T} and experimental work on coherent
atomic recoil lasing \cite{Slama2007aT,Slama2007bT}, to our
knowledge only two recent experimental studies considered in
detail superradiance from ultracold thermal vapors and its
behavior as a function of temperature
\cite{Sadler2007T,Yoshikawa2005T}.

One important feature of superradiance from a bosonic atomic vapor
cooled below the BEC critical temperature is the existence of two
well-separated decay time scales, the fast one attributed to the
Doppler dephasing of the thermal fraction of the gas, and the
other, slower by roughly an order of magnitude, due to the
condensed fraction. These separate time scales were observed by
Yoshikawa and co-workers \cite{Yoshikawa2005T} and exploited by
Sadler \textit{et al.}~\cite{Sadler2007T} to selectively image the
condensed fraction of the vapor over a range of temperatures above
and below the critical temperature. In the latter experiment
\cite{Sadler2007T} superradiance enhanced absorption of light
scattering from a prolate sample of condensed bosonic atoms was
monitored in real time during the superradiance process. A main
goal of the present paper is to provide a theoretical description
of superradiance in ultracold bosonic gases to explain the key
temperature dependent aspects of the above-mentioned experiments
\cite{Yoshikawa2005T,Sadler2007T}, that enable its use as a probe
of coherence in Bose condensates. As such, the present paper is an
extension to finite temperatures of our previous study
\cite{Uys2007T} of the spatial inhomogeneity observed in
absorption images of the probe light at $T=0$.

The paper is organized as follows.  Section II discusses our
theoretical model of a non-interacting, ultra-cold atomic gas
undergoing Raman transitions in a spherical harmonic trap. Section
III presents numerical results demonstrating the existence of two
decay times in the superradiant signal. Section IV shows that in
that regime where the superradiance time scale becomes comparable
to a trap frequency, superradiant scattering occurs predominantly
from atoms in the low-lying energy states of the trap, hence it
may be used to selectively probe those states. We also remark that
superradiance may leave the `cold' and `hot' atoms in two different
electronic states. Finally, section V is a summary and conclusion.

\section{Finite temperature theory}

\begin{figure*}
\includegraphics[angle=0, scale=0.5]{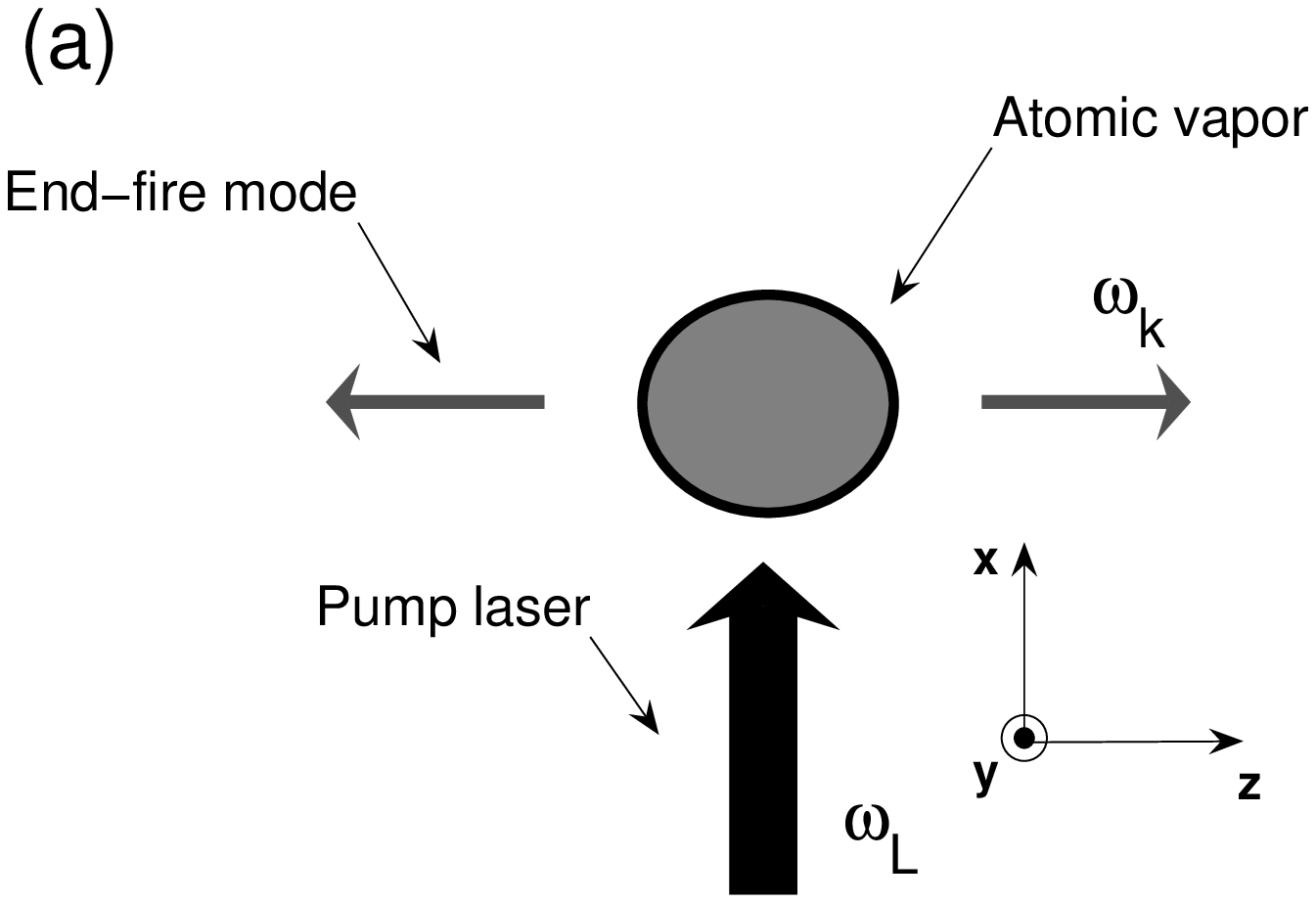}
\includegraphics[angle=0, scale=0.5]{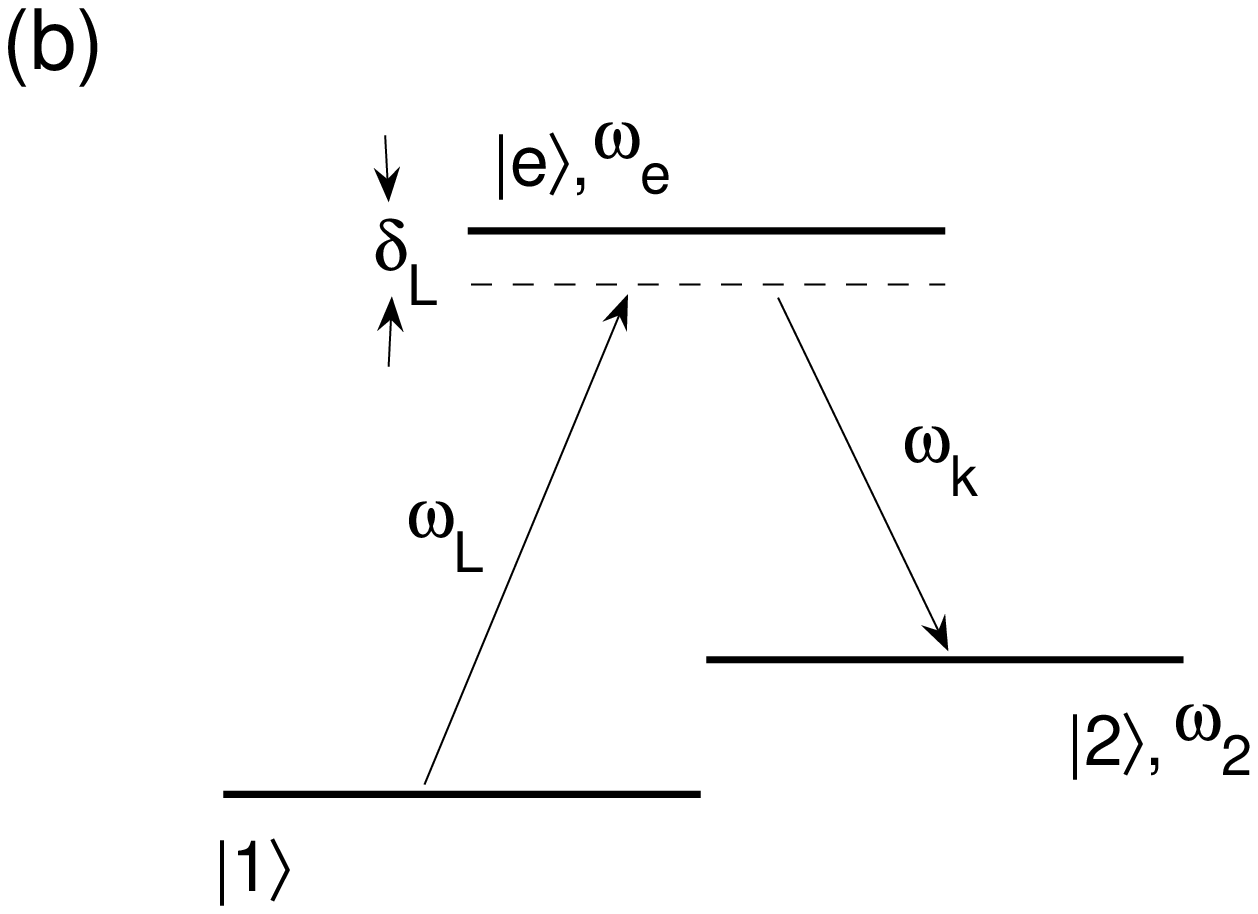}
\caption{(a) Experimental setup - a spherically shaped atomic vapor
is driven by a laser beam incident along the $x$-axis.  The
end-fire modes travelling along the $z$-axis are superradiantly
amplified. The pump light is assumed to drive a $\Lambda$-type
Raman transition for which $\omega_L\gg\omega_2\gtrsim\delta_L$,
see Fig. 1 (b)}\label{expsetupT}
\end{figure*}
We consider $N$ non-interacting ultracold atoms in a spherical
harmonic trap, Fig.~1(a). The atoms undergo $\Lambda$-type Raman
scattering between two electronic ground states $|1\rangle$ and $|2\rangle$
via an excited state $|e\rangle$, see Fig. 1(b). We assume that
the transition $|1\rangle \rightarrow |e\rangle$ is driven by an
off-resonant classical pump laser ${\bf E}_L(t)$ of frequency
$\omega_L$ propagating along the $x$-axis of the trap. For a
spherical trap there are no geometrical effects \cite{Moore1999bT}
that lead to the selection of particular superradiant modes as
would be the case for a cigar-shaped trap, but a preferential
direction can be selected by placing the sample in a low finesse
ring cavity or by first applying a Bragg pulse to generate a seed
matter wave \cite{Kozuma1999T,Inouye1999bT}.  We assume that this is
the case here and that as a result the dominant superradiant modes
propagate along the $z$-axis. We refer to them as end-fire modes
in analogy to the case of elongated samples. The polarization of
the pump light is also chosen parallel to the $z$-axis.  For that
polarization Rayleigh scattering into the end-fire modes is
suppressed due to the angular radiation pattern of Rayleigh
scattering. It does however allow Raman scattering into circularly
polarized end-fire modes which we describe quantum-mechanically in
terms of bosonic creation and annihilation operators
$\hat{a}^\dagger_{\mb{k}}$ and $\hat{a}_{\mb{k}}$, respectively.
The total electric field is then
\begin{eqnarray}
\nonumber\mathbf{\hat E} &=& \hat z\left[E_L
e^{i(\mb{k}_L\cdot\mb{x}-\omega_Lt)} + E_L^*
e^{-i(\mb{k}_L\cdot\mb{x}-\omega_Lt)}\right]\\
\nonumber &+& \mb{\hat
\epsilon_\sigma}\sum\limits_{k}\left[\left(\frac{\hbar\omega_k}{2\epsilon_0
V}\right)^{\frac{1}{2}} \hat{a}_{\mb{k}}(t)e^{i\mb{k\cdot r}} +
h.c.\right],
\end{eqnarray}
where the incident laser field envelope $E_L$ is taken as
constant in amplitude. In terms of the detuning
    \begin{equation}
    \delta_L = \omega_e - \omega_L
    \end{equation}
we have
    \begin{equation}
    \omega_k = \omega_e-\omega_2-\delta_L.
    \end{equation}

We proceed by introducing bosonic matter-field creation and
annihilation operators $\hat{\psi}^\dagger_{i}(\mb{x},t)$ and
$\hat{\psi}_{i}(\mb{x},t)$, that create and annihilate, respectively, an atom
at time $t$ and position $\mb{x}$ in electronic state $|i\rangle =
|1\rangle$, $|e\rangle$ or $|2\rangle$, with
\begin{equation}
\left[\hat{\psi}_{i}(\mb{x},t),\hat{\psi}^\dagger_{j}(
\mathbf{x^\prime},t)\right] =
\delta_{ij}\delta(\mb{x}-\mathbf{x^\prime}).
\end{equation}
Taking $\omega_1 = 0$ the Hamiltonian of the atom-field system is $\hat H = \hat
H_0 +
\hat H_c$, with
\begin{eqnarray}
\label{HfreeqT} \hat H_0 &=& \sum\limits_{\mb k}\hbar \omega_k
\hat{a}_{\mb k}^\dagger\hat{a}_{\mb k} \nonumber
\\
&+&\int
d\mb{x}\left\{\hbar\omega_e\hat{\psi}^\dagger_{e}(\mb{x})\hat{\psi}_{e}(\mb{x})+
\hbar\omega_2\hat{\psi}^\dagger_{2}(\mb{x})\hat{\psi}_{2}(\mb{x})
\right\}\nonumber \\
&+&\sum\limits_{i=1,2,e}\int d\mb{x}\, \hat\psi^\dagger_i(\mb
x)\left[-\frac{\hbar^2\mb\nabla^2}{2m} +V(\mb
x)\right]\hat\psi_i(\mb x),
\end{eqnarray}
while
%\begin{widetext}
\begin{equation}
\label{HcoT} \hat H_c = -\int d\mb{x} \left\{\mathbf{\hat E}\cdot
\mathbf{d}\left[
\hat{\psi}^\dagger_{e}(\mb{x})\hat{\psi}_{1}(\mb{x})+
\hat{\psi}^\dagger_{e}(\mb{x})\hat{\psi}_{2}(\mb{x}) \right]
+h.c.\right\}
\end{equation}
%\end{widetext}
describes the electric dipole interaction between the atoms and
the electromagnetic field, ${\bf d}$ being the dipole moment which
we take to have the same magnitude for both transitions.

For large enough detunings the excited state is not significantly
occupied and may be adiabatically eliminated.  Introducing slowly
varying operators $\tilde\psi_i=\hat\psi_ie^{i\Omega_it}$ and
$\tilde a_\mb{k}=\hat a_\mb{k}e^{i\omega_\mb{k}t}$ and using the
rotating wave approximation yields then the effective interaction
Hamiltonian
    \bwt
    \be
    H_{\rm eff} = -\int\frac{\hbar}{\delta_L} d\mb{x}
    \left\{\Omega_L^2\hat{\psi}^\dagger_{1}(\mb{x})\hat{\psi}_{1}(\mb{x})
    + \Omega_k^2\hat{\psi}^\dagger_{2}(\mb{x})\hat{\psi}_{2}(\mb{x})\hat{a}_{\mb
    k}^\dagger\hat{a}_{\mb k}+\Omega_L\Omega_k\left(\hat{\psi}^\dagger_{2}(\mb{x})\hat{\psi}_{1}(\mb{x})\tilde
    a_\mb{k} e^{i(\mb{k}_L-\mb{k})\cdot\mb{x}}+h.c.\right)
    \right\}.
    \label{HeffT}
    \ee
    \ewt
Here $\Omega_L=dE_L/\hbar$ is the Rabi frequency of the incident
field and $\Omega_k=d\sqrt{\omega_k/(2\epsilon_0\hbar V)}$.

Expanding the atomic field operators in states of the trap,
    \be
    \hat\psi_i(\mb x)=\sum\limits_{i=0}^\infty
    \phi_n(\mb x)\hat c_{ni} \label{oscexpT}
    \ee
where $H_0\phi_n(\mb x) =  \hbar\nu_n\phi_n(\mb x)$ and the
subscript $n$ labels generically excitations in all three
dimensions of the trap, $n = \{n_x, n_y, n_z\}$, and
$\nu_n=(n_x+n_y+n_z+3/2)\omega_t$, substituting the expansion
(\ref{oscexpT}) into Hamiltonians (\ref{HeffT}) and
(\ref{HfreeqT}), and denoting the expectation values
$\langle\tilde a_\mb{k}\rangle = a_\mb{k}$, we obtain the
Heisenberg equations of motion
\begin{widetext}
\begin{eqnarray}
\frac{\partial \langle  \tilde c_{j1}^\dagger \tilde c_{n1}\rangle}
{\partial\tau}&=& i\sum_{\mb{k} m}\left[\eta_{nm}\left(-\mb{q}\right)
a^\dagger_\mb{k} \langle \tilde c_{j1}^\dagger \tilde c_{m2}\rangle
e^{i(\nu^\prime_n-\nu^\prime_m)\tau}-\eta_{jm}^*\left(-\mb{q}\right)
a_\mb{k} \langle \tilde c_{m2}^\dagger \tilde c_{n1}\rangle
e^{i(\nu^\prime_m-\nu^\prime_j)\tau}\right],\label{density1T}\\
\frac{\partial  \langle \tilde c_{j2}^\dagger \tilde c_{n2}\rangle}
{\partial\tau}&=& i\sum_{\mb{k} m}\left[\eta_{nm}\left(\mb{q}\right)
a^\dagger_\mb{k}\langle  \tilde c_{j2}^\dagger \tilde c_{m1}\rangle
e^{i(\nu^\prime_n-\nu^\prime_m)\tau}-\eta_{jm}^*\left(\mb{q}\right)
a_\mb{k}\langle  \tilde c_{m1}^\dagger \tilde c_{n2}\rangle
e^{i(\nu^\prime_m-\nu^\prime_j)\tau}\right],\label{density2T}\\
\frac{\partial \langle  \tilde c_{j2}^\dagger \tilde
c_{n1}\rangle} {\partial\tau}&=&
i\left[(\frac{\Omega_L}{\Omega_k}-\sum_\mb{k}
\frac{\Omega_k}{\Omega_L} a^\dagger_\mb{k}  a_\mb{k})\right]
\langle \tilde c_{j2}^\dagger \tilde c_{n2}\rangle +i\sum_{\mb{k}
m} \left[\eta_{nm}\left(-\mb{q}\right) a^\dagger_\mb{k}\langle
\tilde c_{j2}^\dagger \tilde c_{m2}\rangle e^{i(\nu^\prime_n-
\nu^\prime_m)\tau}-\eta_{jm}^*\left(\mb{q}\right) a_\mb{k}\langle
\tilde c_{m1}^\dagger \tilde c_{n1}\rangle e^{i(\nu^\prime_m
-\nu^\prime_j)\tau}\right].\nonumber \\ \label{polarizationT}
\end{eqnarray}
\end{widetext}

Here the dimensionless time $\tau = \Omega_Rt$ where $\Omega_R =
\Omega_L\Omega_k/\delta_L$ is an effective `two-photon' Rabi
frequency, $\nu_n^\prime = \nu_n/\Omega_R$, the slowly varying
operators $\tilde c_{ni} = \hat c_{ni}e^{i\nu^\prime_nt}$ and the
overlap function
    \be
    \eta_{nj}(\mb{q})=\int
    \phi^*_n(\mb{x})\phi_j(\mb{x})e^{i\mb{q}\cdot \mb{x}}d\mb{x},
    \label{overlapT}
    \ee
    where the recoil momentum is $\mb{q}=\mb{k}_L-\mb{k}$.  A grayscale rendering of the absolute value of
$\eta_{nj}(\mb{q})$ is plotted for the first 25 trap levels and the trap and light field parameters introduced in
section III. 
\begin{figure}
\includegraphics[angle=0, scale=0.5]{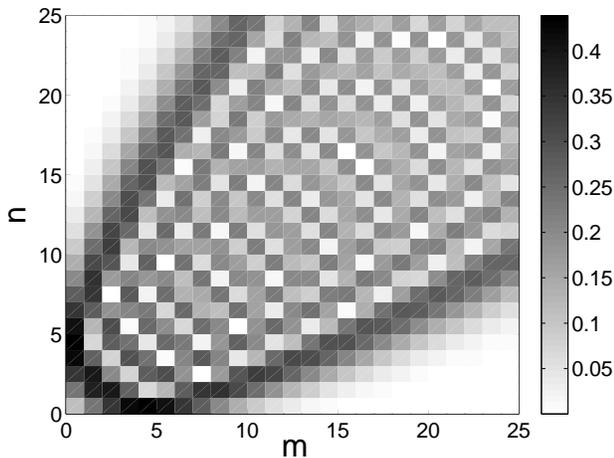}
\caption{Gray scale rendering of the absolute value of the overlap
function $\left|\eta_{mn}(k)\right|$ for the first 25 trap
levels.  Here we've assumed a trap frequency
$\omega_t = 2\pi\cdot800$ rad/s and $k_L=2\pi/795$ nm 
corresponding to the $F=1\rightarrow F^\prime=1$ transition of $^{87}$Rb.}\label{frankcfigT}
\end{figure}
Finally, the time evolution of the end-fire modes is given by
\begin{eqnarray}
\frac{\partial  a_\mb{k}}{\partial\tau} &=&
i\Omega_k/\Omega_L\sum\limits_m\langle\tilde c_{m2}^\dagger\tilde
c_{m2}\rangle a_\mb{k} \nonumber \\
&+& i\sum_{mn}\eta_{mn}\left(\mb{q}\right) \langle \tilde
c_{m2}^\dagger\tilde c_{n1}\rangle
e^{i(\nu^\prime_m-\nu^\prime_n)t} - \Gamma a_\mb{k},
\label{efmeqmT}
\end{eqnarray}
where the last term is a phenomenological decay added to account
for the escape of the photons from the sample. This term should in
principle be accompanied by appropriate quantum noise operators to
guarantee that the commutation relations of the operators $a_{\bf
k}$ are preserved at all times. These noise operators are also
important in that they trigger the superradiant amplification and
determine the field fluctuations in its early stages \cite{Gronchi1978, Haake1979}. This paper concentrates however on
the later,
classical stages of the process, and we follow the standard
procedure of introducing a small classical seed for the initial
amplitudes of the end-fire modes, which are then described as
classical fields. At that level of approximation the
phenomenological decay term in Eq. (\ref{efmeqmT}) is
appropriate. Note also that by factorizing the expectation values
of the optical and matter-wave fields, we have neglected any
quantum correlation and entanglement that may build up between the
optical and matter-wave fields fields.

The atoms obey initially an equilibrium Bose distribution. We
determine the distribution for excitations in the $z$-direction by
averaging over the $x$ and $y$-directions,
    \be
    \langle c^\dagger_{n1}c_{n1}\rangle = \frac{1}{\mathcal
    N}\sum\limits_{n_x,n_y=0}^\infty\frac{1}{e^{\hbar\omega_t\left(n_x
    + n_y+ n_z\right)/k_BT}-1},\label{bosedistT}
    \ee
where $\mathcal N$ is a normalization factor chosen to fix the
thermal fraction of atoms, $n_{\rm th}=\sum\limits_{n=0}^\infty
\langle c^\dagger_{n1}c_{n1}\rangle$, at the value  $n_{\rm th} =
N(T/T_c)^3$ determined by the temperature of the sample
\cite{Pitaevskii2003T}, where $T_c$ is the critical temperature.

We conclude this section by remarking that in order to keep the
computations manageable we have assumed that the light field
intensity is uniform within the sample, thereby neglecting spatial
effects that have previously been shown to play an
important role in the growth of the end-fire modes \cite{Zobay2005T,Zobay2006T,
Uys2007T}. The role of these effects at finite temperature remains
therefore an open question. In the following
simulations we also assume for simplicity that the atoms recoil
only along the $z$-axis and we are therefore only interested in
excitations along that direction.

\section{Dephasing}

\begin{figure}
\includegraphics[angle=0, scale=0.45]{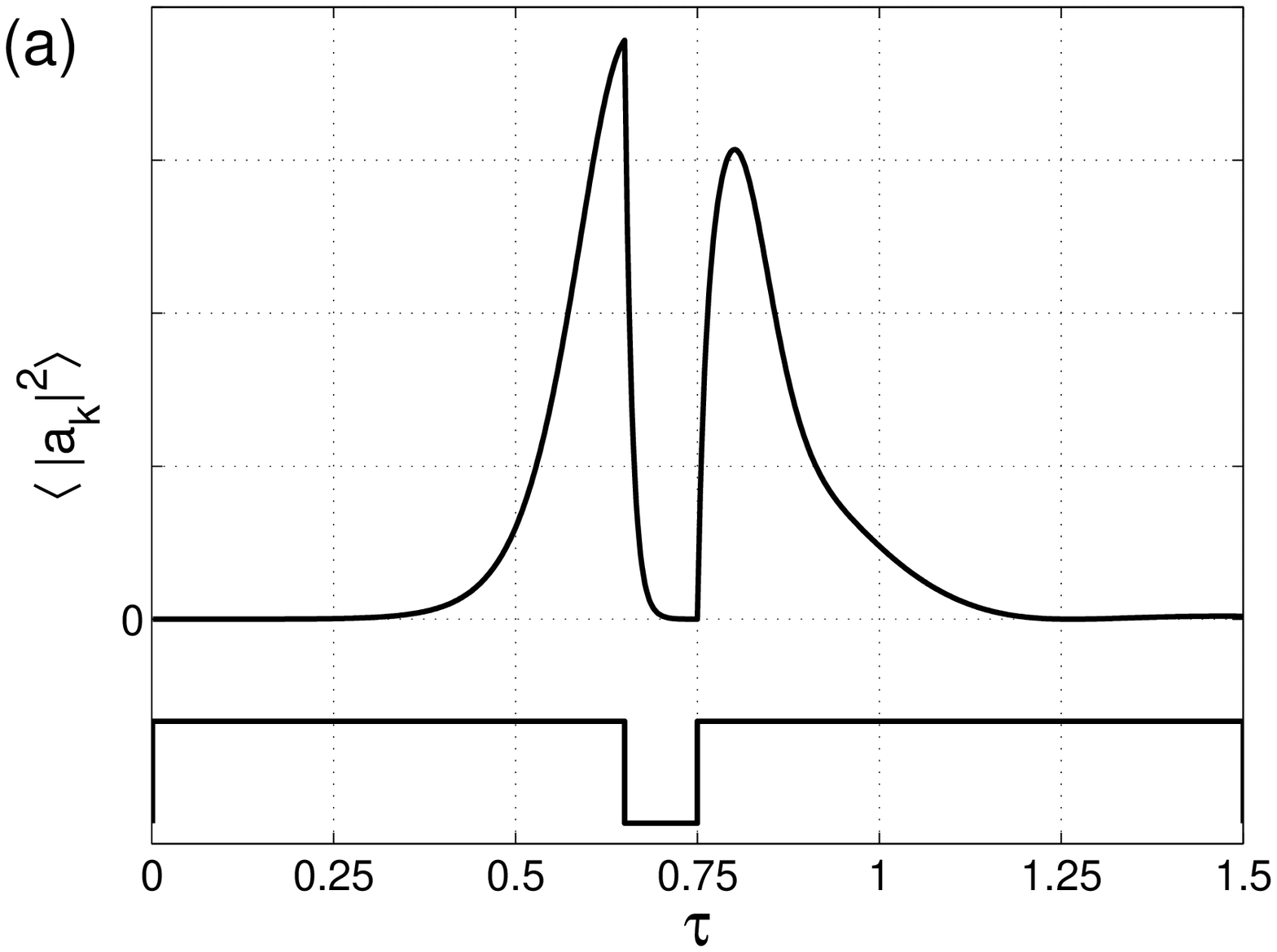}
\includegraphics[angle=0, scale=0.45]{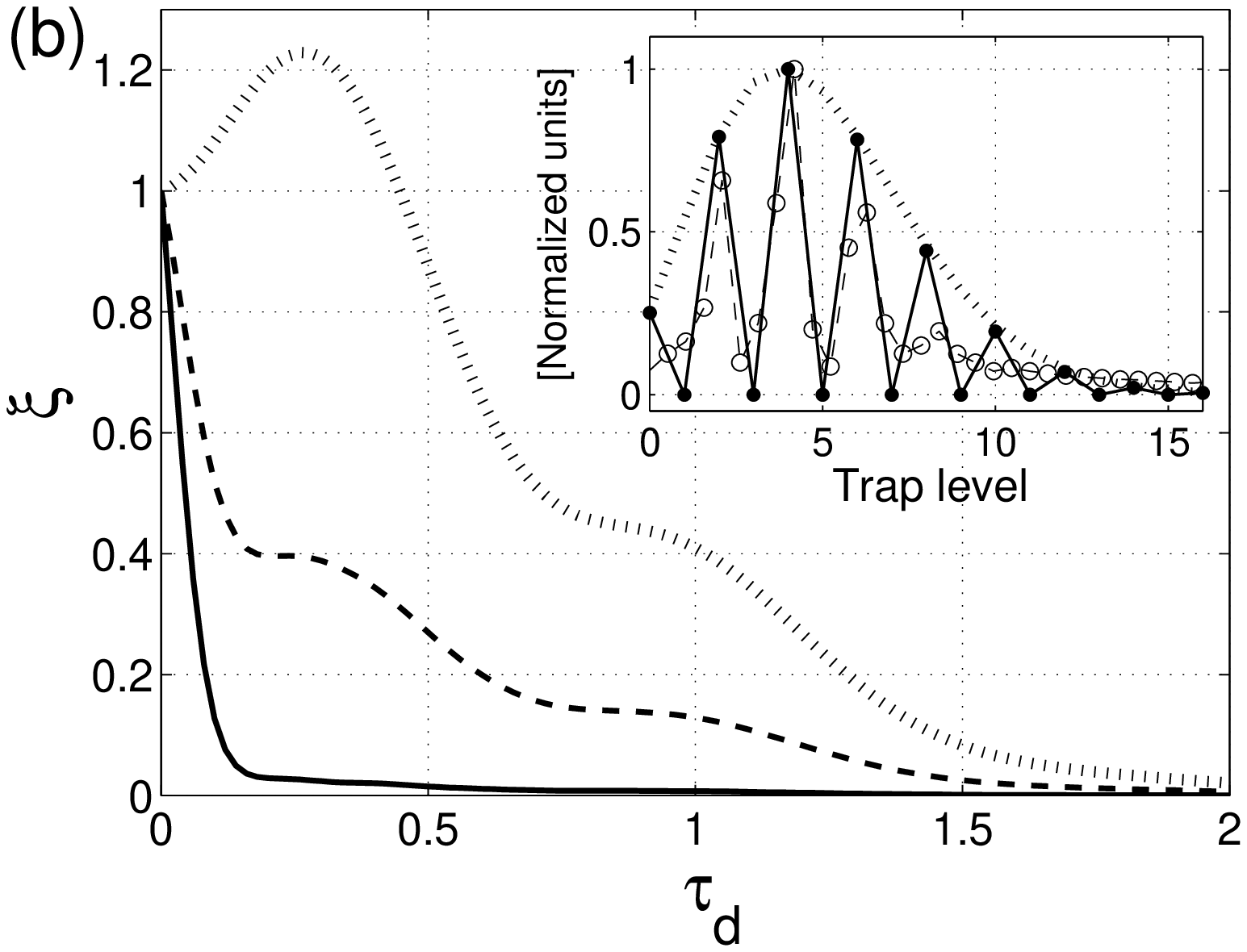}
\caption{(a) Pump-probe spectroscopy.  The lower curve shows the
pump-probe sequence and the upper curve the resulting end-fire
mode intensity. The ratio of the peak heights directly after to
directly before the delay is used as a measure of coherence. (b)
Dephasing. Dotted line: post- to pre-delay intensity signal ratio
for a vapor with $0.95$ condensed fraction. For a condensed
fraction of only $0.1$ (solid line) the decay is bimodal,  the
initial rapid decay being attributed to the thermal fraction.
Dashed line: intermediate case, $n_c=0.5$. (Inset) Solid circles:
polarization $\left|\langle\tilde c_{m2}^\dagger\tilde
c_{01}\rangle\right|$; open circles Fourier components of the
signal for $n_c=0.95$ in the main figure. The dotted line envelope
shows the overlap function $\left|\eta_{m0}(\mb{q})\right|$.
}\label{decayfigT}
\end{figure}
\begin{figure*}
\includegraphics[angle=0, scale=0.3]{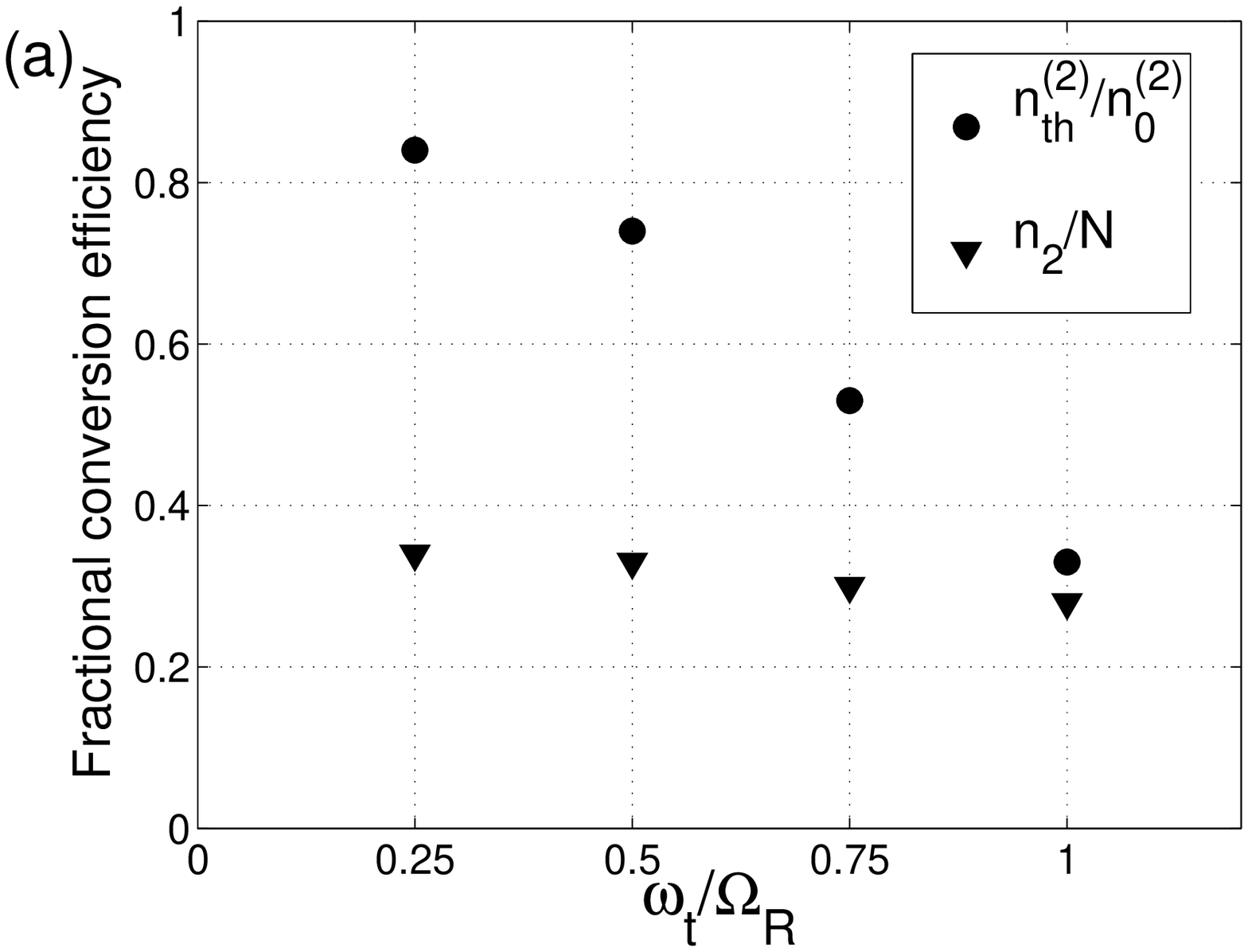}
\includegraphics[angle=0, scale=0.3]{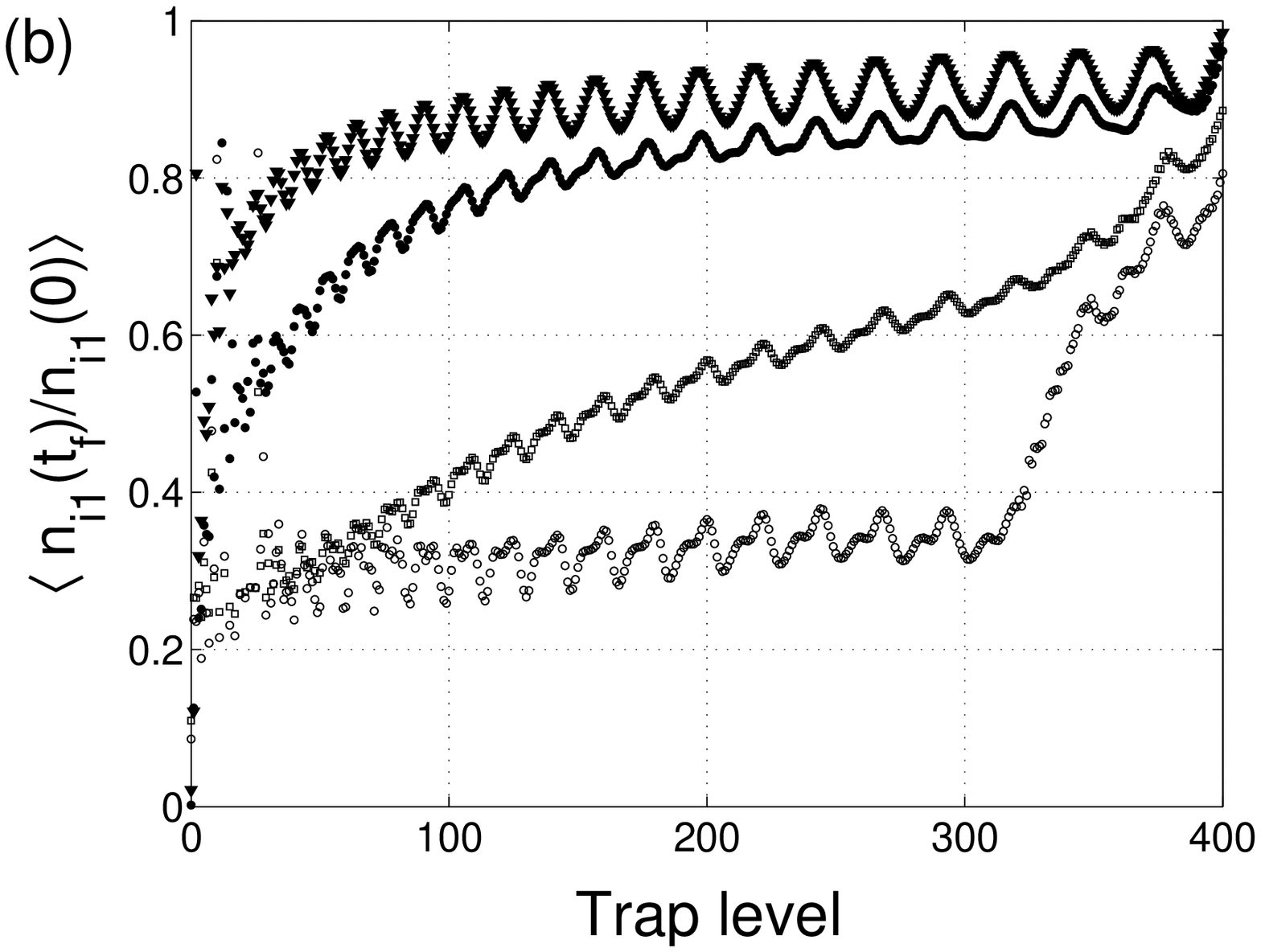}
\includegraphics[angle=0, scale=0.3]{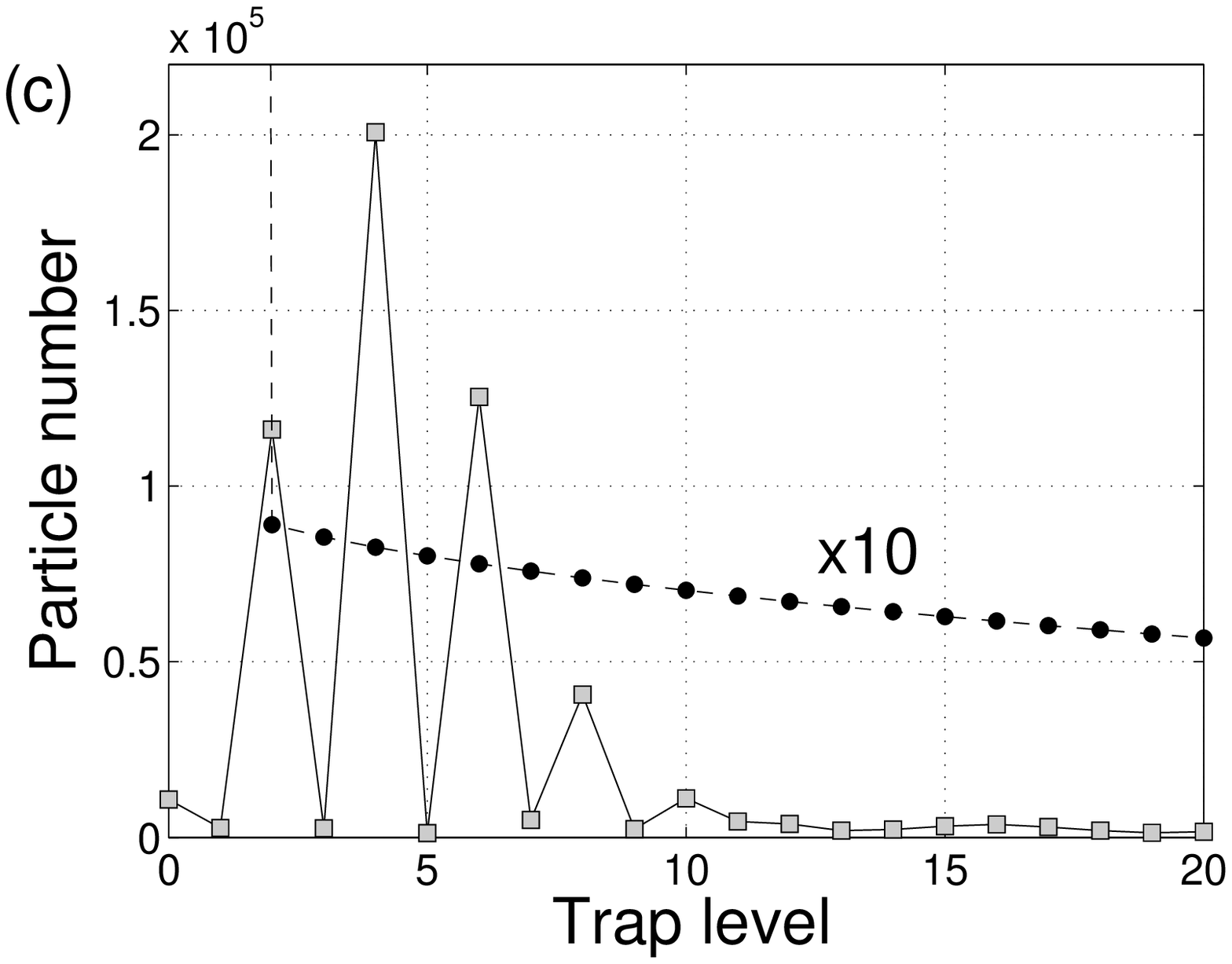}
\caption{Summary of numerical simulations for $n_c=0.5$. (a)
Circles: Ratio of the total number of thermal atoms $(n^{(2)}_{\rm
th})$ to total number of condensed atoms $(n^{(2)}_c)$ transferred
from the state $|1\rangle$ at peak superradiant intensity ;
triangles: fraction of total number of atoms scattered to the
state $|2\rangle$ (condensed and thermal included), as a function
of $\omega_t/\Omega_R$.  (b) Fraction of atoms left in ground
state $|1\rangle$ at the end of the pulse complete by trap level
for $\omega_t/\Omega_R = 0.25$, $0.5$, $0.75$ and $1.0$. (c)
Initial distribution of atoms by trap level (solid circles) and
distribution of atoms in state $|2\rangle$ at the end of the pulse
for $\omega_t/\Omega_R = 1.0$. The initial distribution was multiplied by a factor of $10$ for
visibility.}\label{cohimagfigT}
\end{figure*}

This section summarizes numerical results that illustrate the
dynamics of the thermal and condensed fractions of the vapor,
showing that these dynamics are characterized by two distinct time
scales. Specifically, we simulate a pump-probe technique that was
experimentally implemented by Yoshikawa and co-workers
\cite{Yoshikawa2005T} to measure superradiant coherence. This
procedure entails applying a superradiant pump pulse until the
end-fire mode intensity reaches a maximum.  The growth of the end-fire mode occurs in response to the build-up of atomic
polarization (a polarization grating forms), Eq.~(\ref{efmeqmT}).  At that point the
pulse is turned off for a variable delay time, $\tau_d$, during
which the polarization grating deteriorates due to Doppler
dephasing.  The pump beam is then turned back on to its original
intensity to probe the remaining coherence.  Had the polarization
grating remained intact the superradiant intensity would
immediately return to its pre-delay level, however, due to Doppler
dephasing the post-delay intensity is reduced. The ratio of the
pre- and post-delay end-fire mode intensities,
    \be
    \xi = |a_\mb{k}(\tau_{\rm max}+\tau_d)|^2/|a_\mb{k}(\tau_{\rm max})|^2
    \label{signalratioT},
    \ee
is therefore a measure of the decoherence of the superradiant
pulse.

Figure~\ref{decayfigT}(a) illustrates the detector response (upper
curve) to a square pulse pump-probe sequence (lower curve).  Here
the end-fire mode intensity was convoluted with the response
function of the detector, assumed to have a (dimensionless)
response time of $\tau^\prime=0.01$.

Our simulations are for a sample of $N=10^6$ atoms in a
three-dimensional, isotropic harmonic trap with trap frequency
$\omega_t = 2\pi\cdot800$ rad/s and, with reference to \cite{Sadler2007T} we use the $k_L=2\pi/795$ nm 
corresponding to the $F=1\rightarrow F^\prime=1$ transition of $^{87}$Rb.  We use $\omega_t/\Omega_R =0.5$
and included 400 trap levels in the simulations. A tight trap was chosen for computational reasons: For that value of
$\omega_t$ the occupation of levels higher than $n=400$ remains
negligible at all times for the conditions that we consider
\footnote {The relatively high trap frequency that we consider is
guided solely by numerical considerations: an atom in the lowest
trap level that experiences a recoil $\hbar k_L=2\pi\hbar/795$ nm along either of
the axes is excited to a trap level $n_r \approx \hbar
k^2/2M\omega_{t}$, or $n_r\approx 10^3$ for typical trap
frequencies ($\omega_t = 10\pi$ rad/s in \cite{Sadler2007T}). Our
choice of $\omega_t$ reduces $n_r$  by two orders of magnitude,
rendering the problem computationally significantly more
tractable.}. We choose $\Omega_L/\Omega_k = 1000$ and neglect
laser pump depletion. Since the decay of the light field due to
its escape from the atomic sample, see Eq.~(\ref{efmeqmT}), is the
fastest process by several orders of magnitude the state of the
scattered light field is nearly instantaneously determined by the
state of the atomic fields, so that
    \be
    a_\mb{k} \approx \frac{i}{\Gamma}\sum_{mn}\eta_{mn}
    \left(\mb{q}\right) \langle \tilde c_{m2}^\dagger\tilde c_{n1}\rangle
    e^{i(\nu^\prime_m-\nu^\prime_n)t}.
    \label{akapproxT}
    \ee

Either a seed atomic polarization or a seed end-fire mode
occupation is required to initiate the superradiant growth. In
this paper we usually assume that the end-fire mode has an initial
value $a_\mb{k}=\sqrt{10}$ and that all atoms are in the state
$|1\rangle$ obeying the distribution~(\ref{bosedistT}). The first
numerical iteration of Eqs.~(\ref{density1T})-(\ref{polarizationT})
then creates a small initial polarization.  Thereafter we use
Eq.~(\ref{akapproxT}) to determine the instantaneous value of the
end-fire mode, with $\Gamma\approx(10^4\sim10^5)\times\Omega_R$.

Figure \ref{decayfigT}(b) plots $\xi(\tau_d)$, the post- to
pre-delay intensity ratio,  Eq.~(\ref{signalratioT}), as a
function of the pump-probe delay time for three different sample
temperatures. Since the superradiance process is initiated by
quantum fluctuations leading to spontaneous emission of photons
into the end-fire modes, shot-to-shot fluctuations occur in the
relative amplitudes and phases of the left- vs. right-propagating
end-fire modes \cite{Uys2007T}. To account for these fluctuations
we chose, for the purposes of Fig.~\ref{decayfigT}(b), the initial
seed amplitudes from a random Gaussian distribution centered
around $a_\mb{k}=\sqrt{10}$ with a standard deviation of
$\sigma=\sqrt{10}/2$ and with a random overall phase.  Each curve
in Fig.~\ref{decayfigT}(b) represents an average over fifty runs
of pump-probe experiments.

The dotted line corresponds to a nearly pure
condensate, with a condensed fraction $n_c=0.95$. After an initial
increase for short times due to the continued build-up of the
atomic polarization , the signal ratio decreases to half the initial intensity with
a decay time of $\tau_c\approx 0.3$.

The slow oscillations in $\xi(\tau_d)$ can be understood by
comparing the Fourier transform in one single shot of the
experiment to the components of the atomic polarization $|\langle
c_{m2}^\dagger c_{01} \rangle |$ between the trap ground state and
its excited levels, see insert of Fig.~\ref{decayfigT}(b). As
expected, there is a one-to-one correspondence between these
frequencies. Hence the oscillations in $\xi(\tau_d)$ are a
signature of the specific trap characteristics. In the inset the
atomic coherence is only significant between even trap states.
This is a combined consequence of the two end-fire mode amplitudes
and phases having been chosen equal in this shot, and of the form
of the overlap function $(\eta_{j0}(q))$ between the lowest trap
level and higher levels, which are alternately purely real and
purely imaginary, due to the harmonic oscillator states being
alternately even and odd. In Eq.~(\ref{polarizationT}), only the
last term on the right-hand side contributes initially.  Then, by
choosing the initial overall phase of the left-propagating
end-fire mode to be the same as that of the right-propagating one,
the left and right contributions for all odd trap levels cancel
out since the overlap functions for those modes are purely
imaginary. The cancellation is no longer exact if either the
phases of the left and right-propagating end-fire modes are not
the same or if their initial amplitudes are not equal, as is in
general the case due to fluctuations.

As the temperature approaches $T_c$, $\xi(\tau_d)$ undergoes a
rapid initial decay on a timescale $\tau_{\rm th}$ roughly an
order of magnitude faster than the slow decay characterized by
$\tau_c$. This is illustrated for $n_c=0.1$ as the solid line in
Fig.~\ref{decayfigT}(b). These two time scales were experimentally
observed by Yoshikawa \textit{et al.}, see Fig.~3 in
Ref.~\cite{Yoshikawa2005T}.  The appearance of a slow decay below
the critical temperature $T_c$, combined with the disappearance of
the rapid decay at $T \rightarrow 0$, clearly points to the fact
that they are associated with the condensed and thermal fractions
respectively. The existence of these two time scales was exploited
by Sadler \textit{et al.} \cite{Sadler2007T} to image selectively
the condensed fraction of an ultra-cold Bose gas below the
critical temperature while remaining blind to the thermal
fraction, as we discuss next.

\section{Thermally selective scattering}

This section further discusses how the Doppler broadening of the
thermal fraction inhibits superradiance when the effective Rabi
frequency $\Omega_R$ becomes comparable to the trap frequency,
thereby making superradiance a sensitive probe of the condensed
fraction of the sample \cite{Sadler2007T}.

The circles in Fig.~\ref{cohimagfigT}(a) show the ratio
$n^{(2)}_{\rm th}/n^{(2)}_{c}$, the number of atoms transferred to
the electronic state $|2\rangle$ from the excited trap levels to
the number of atoms transferred from the trap ground state to
$|2\rangle$, at peak end-fire mode intensity and as a function of
$\omega_t/\Omega_R$ for an initial condensed fraction $n_c=1/2$.
As the ratio $\omega_t/\Omega_R$ increases, fewer thermal atoms
participate in the superradiant scattering as a consequence of
Doppler dephasing. The triangles, which show the total number of
scattered atoms, confirm that indeed, the bulk of the superradiant
emission is associated to the condensed atoms in that case.

This behavior is further illustrated by plotting the fraction of
atoms left in the electronic state $|1\rangle$ at the end of the
superradiant emission as a function of trap level. This is
illustrated in Fig.~\ref{cohimagfigT}(b) for the cases (from
lowest to highest curve) $\omega_t/\Omega_R = 0.25$, $0.5$ and
$0.75$ and $1.0$. When superradiance occurs on a time scale faster
than the inverse trap frequency, all levels contribute roughly in
proportion to their initial occupation \footnote{The rapid rise at
the tail-end of that curve is a consequence of the finite number
of trap levels in the simulations. Since the population of high
levels remains low at all times this artifact does not
significantly affect our conclusions.}. But as $\omega_t/\Omega_R$
becomes of order unity, the lower trap levels contribute
significantly more in proportion to their initial occupation than
the higher levels which Doppler dephase faster.

The selective scattering of atoms initially in the deepest trap
levels implies that the scattered atoms are in effect colder
than the unscattered atoms --- a situation somewhat akin to
evaporative cooling, except that in the present case it is the
warmer atoms that are ``left behind.'' The squares in
Fig.~\ref{cohimagfigT}(c) show the atomic distribution of atoms in
electronic state $|2\rangle$ at the end of the superradiant
emission for the first 20 trap levels for $n_c=0.5$,
$\omega_t/\Omega_R=1.0$ and $N=10^6$.  The solid circles show the
initial distribution of thermal atoms in state $|1\rangle$ (the
occupation number of the lowest trap level is off scale and is not
shown here and the initial distribution was multiplied by a factor of 10 for visibility).

We use the root-mean-square deviation of the center-of-mass energy
of the atoms as a measure of their temperature, and compare its
initial value $\Delta\epsilon_1$ for atoms in the state
$|1\rangle$, to that of the atoms in the electronic ground state
$|2\rangle$, $\Delta\epsilon_2$, at the end of the superradiant
emission. For $\omega_t/\Omega_R = 1.0$ we find $\Delta\epsilon_2
= \Delta\epsilon_1/2$, a modest reduction in rms energy. This is
despite the fact that several trap levels wind up being
significantly populated, as opposed to only the lowest trap level
being significantly populated initially and follows from the
comparatively few thermal atoms undergoing Raman scattering, as we
have seen. For the case at hand, 58\% of the particles in the
state $|1\rangle$ initially occupy the first 10 trap levels, but
after scattering 82\% of the particles scattered to $|2\rangle$
occupy these 10 levels.

This effect can be enhanced if the trap frequency is chosen so as
to approximately match the photon recoil frequency, $\hbar
q^2/2M\approx \omega_t$, the spatial width of the ground state of
the atom trap thus being of the same order as the wavelength of
light.  This implies that the Lamb-Dicke parameter, $\eta=
k_L\sqrt{\hbar/2m \omega_t}$, which serves as a measure of the
amount of coupling between the motional and internal states of the
atoms, is of order $\eta \simeq 1$. The overlap function
associated with the ground state, $\eta_{0m}(\mb{q})$, is then
appreciable only for $m=1$, and the condensed atoms will only
recoil to that level, instead of several levels as was the case in
Fig.~\ref{cohimagfigT}(c).

\section{Conclusion}

In this paper we have explored theoretically various aspects of
superradiance from ultracold, but finite temperature atomic gases.
In particular we've illustrated explicitly the existence of two
well separated coherence timescales associated with the condensed
and thermal fractions of the gas respectively. In the regime where
Doppler dephasing plays an important role, we found that
superradiant scattering takes place predominantly from atoms in
the lowest lying trap levels as these are less prone to the
Doppler dephasing.  That effect was exploited in
\cite{Sadler2007T} to image selectively the condensed fraction of
an ultra-cold gas.  We demonstrated that as a consequence atoms in
the scattered state have a reduced rms deviation from the average
trap energy as compared to the atoms in the initial state, and are
in that sense colder than atoms in the initial state.

Our calculations considered a system with end-fire modes recoiling
perpendicularly to the pump beam. In a system where the pump beam
propagates collinearly to the scattered beams, such as an
elongated condensate pumped along the long axis, it is possible to
suppress Raman scattering into the backward end-fire mode, as
compared to the forward end-fire mode, by exploiting the same
dephasing effects that lead to thermally selective scattering.
This results because the recoil momentum associated with the
forward end-fire mode is small compared to that due to backward
scattering.  Hence it is possible to realize conditions such that
the forward scattering Lamb-Dicke parameter $\eta_f \ll 1$ and the
backward scattering one $\eta_b \gg 1$, implying negligible coupling between the ground state and higher motional
states for forward scattering, but strong coupling and accompanying dephasing for the backward scattering case. This
effect would not be present in Rayleigh scattering experiments using the same
geometry, since there the pump beam simply sees a phase shift in
the forward direction and only the backward mode is superradiantly
enhanced.

\section*{Acknowledgments}

It is a pleasure to thank Mishkat Battacharya, Wenzhou Chen,
Omjyoti Dutta and Swati Singh for fruitful discussions.

This work is supported in part by the US Army Research Office, the
National Science Foundation and the US Office of Naval Research.

%\bibliography{superT}

\end{document}